\documentclass[openacc]{rstransa}

\usepackage[compress,comma,square]{natbib}

\setcitestyle{numbers}
\usepackage{float}
\usepackage[caption=false]{subfig}

\begin{document}

\title{Molecular dynamics lattice gas equilibrium distribution function for Lennard\nobreakdash-Jones particles}

\author{
Aleksandra Pachalieva$^{1, 2}$ and Alexander J. Wagner$^{3}$}

\address{$^{1}$Center for Nonlinear Studies, Los Alamos National Laboratory, Los Alamos, NM 87545, USA \\
$^{2}$Department of Mechanical Engineering, Technical University of Munich, 85748 Garching, Germany\\
$^{3}$Department of Physics, North Dakota State University, Fargo, ND 58108, USA}

\subject{computational physics, coarse\nobreakdash-graining, fluid mechanics}

\keywords{molecular dynamics, lattice gas method, lattice Boltzmann method, coarse-graining}

\corres{Alexander J. Wagner\\
\email{alexander.wagner@ndsu.edu}}

\begin{abstract}
The molecular dynamics lattice gas method maps a molecular dynamics simulation onto a lattice gas using a coarse\nobreakdash-graining procedure. This is a novel fundamental approach to derive the lattice Boltzmann method by taking a Boltzmann average over the molecular dynamics lattice gas. A key property of the lattice Boltzmann method is the equilibrium distribution function, which was originally derived by assuming that the particle displacements in the molecular dynamics simulation are Boltzmann distributed. However, we recently discovered that a single Gaussian distribution function is not sufficient to describe the particle displacements in a broad transition regime between free particles and particles undergoing many collisions in one time step. In a recent publication, we proposed a Poisson weighted sum of Gaussians which shows better agreement with the molecular dynamics data. \textcolor{black}{We derive a lattice Boltzmann equilibrium distribution function from the Poisson weighted sum of Gaussians model and compare it to a measured equilibrium distribution function from molecular dynamics data and to an analytical approximation of the equilibrium distribution function from a single Gaussian probability distribution function.}
\end{abstract}

\begin{fmtext}
\end{fmtext} 
\maketitle
\section{Introduction}
The molecular dynamics lattice gas (MDLG) method \cite{parsa_lattice_2017, parsa2020} uses a coarse\nobreakdash-graining procedure to establish a direct link between microscopic methods -- in particular, molecular dynamics (MD) simulation, and mesoscale methods such as lattice gas (LG) \cite{frisch_lattice-gas_1986-1,blommel_integer_2018}, and lattice Boltzmann methods (LBM) \cite{qian_lattice_1992, he1997theory}. The MDLG fully relies on MD data and as such it rigorously recovers the hydrodynamics of the underlying physical system, and can be used to verify the behavior and examine the properties of the LG or the LBM methods directly without using the standard kinetic theory approach. Aspects that can be examined include fluctuating \cite{Ladd1993,Adhikari_2005,dunweg2007,strand2016}, thermal \cite{he1998novel,mcnamara1995stabilization}, multi\nobreakdash-phase and multi component systems \cite{osborn1995lattice, shan1995multicomponent,he1997theory,briant2004lattice}.

A key feature in the LBM is the equilibrium distribution function. The LBM equilibrium distribution was originally derived by analogy to the continuous Boltzmann equation, where the equilibrium distribution for the velocities is a Maxwell Boltzmann distribution. Similarly, the LBM moments of the discrete velocity distribution were matched, to the degree possible, with the velocity moments of the Maxwell Boltzmann distribution. In the alternative derivation of the LBM from MD, it was shown that these previously postulated equilibrium distributions are indeed, at least approximately, consistent with the MDLG approach for specific discretization combinations for lattice and time spacing. 

In the original MDLG calculation of the equilibrium distribution by Parsa \textit{et al.}\,\cite{parsa_lattice_2017}, it was assumed that the particle displacements in the molecular dynamics simulation are also Boltzmann distributed. This assumption gave an adequate prediction of the global equilibrium distribution function of the lattice Boltzmann method.  \textcolor{black}{However, later on by examining more carefully the equilibrium system, we noticed small deviations (up to 5\%) between the analytically predicted and the measured equilibrium distribution functions.} These deviations were traced back to the prediction of the one\nobreakdash-particle displacement distribution function. In Pachalieva \textit{et al.}\,\cite{pachalieva2020}, we proposed a correction of the displacement distribution function, which shows that a dilute gas with area fraction of $\phi = 0.0784$ and temperature of 20 LJ is better approximated by a Poisson weighted sum of Gaussians (WSG) probability distribution function. \textcolor{black}{This probability distribution function takes into account that after a time step $\Delta t$ the particles can be divided into groups depending on the number of collisions they have experienced. In principle, the timing of the collisions should be random (given by a Poisson process), however, the resulting integrals over the collision times do not allow for an analytical solution. Thus, we assume that the particle collisions are evenly spaced, which may introduce a small error but it makes the resulting displacements again Gaussian distributed. For details, please refer to \cite{pachalieva2020}. The Poisson weighted sum of Gaussians probability distribution function also delivers better results for a purely ballistic and purely diffusive regimes (for very small or very large time steps respectively), where the Poisson WSG formulation is reduced to a single Gaussian.} In the current publication, we show that the original premise of the paper \cite{pachalieva2020} does indeed hold. We derive the MDLG equilibrium distribution function from the Poisson WSG one\nobreakdash-particle displacement function and show that it compares favourably to a measured equilibrium distribution function from molecular dynamics (MD) simulation, whereas the single Gaussian equilibrium distribution function is a much poorer prediction. Our findings show that the Poisson WSG approximates the measured equilibrium distribution function significantly better.

The rest of the paper is summarized as follows: \textcolor{black}{We briefly describe the MDLG analysis method in Section \ref{sec:mdlg}. In Section \ref{sec:equilibrium_df}, we derive the equilibrium distribution function from one\nobreakdash-particle displacement function. In Section \ref{sec:equilibrium_df}\,(a), we show how to derive the equilibrium distribution function when the distribution is given by a single Gaussian and in Section \ref{sec:equilibrium_df}\,(b) when the displacements are instead distributed according to a Poisson WSG one\nobreakdash-particle displacement function. In Section \ref{sec:sim_setup}, we give a detailed description of the MD simulation setup used to obtain the MD data. The MD trajectories are later used to validate the theoretical solutions of the equilibrium distribution function. In Section \ref{sec:results}, we compare the equilibrium distribution function obtained on one hand from theory, using either a single Gaussian or the novel Poisson weighted sum of Gaussians probability distribution function, and on the other hand, measured from MD data. Our analysis shows significant improvement of the equilibrium distribution function analytical prediction when the Poisson WSG model is used. Finally, in Section \ref{sec:outlook}, we give a brief conclusion and suggestions for future work .}

\section{Molecular dynamics lattice gas method}
\label{sec:mdlg}
\begin{figure}[!t]
\centering
\subfloat[\label{subfig-1:mdlg_dynamics}]{{\includegraphics[width=0.45\textwidth]{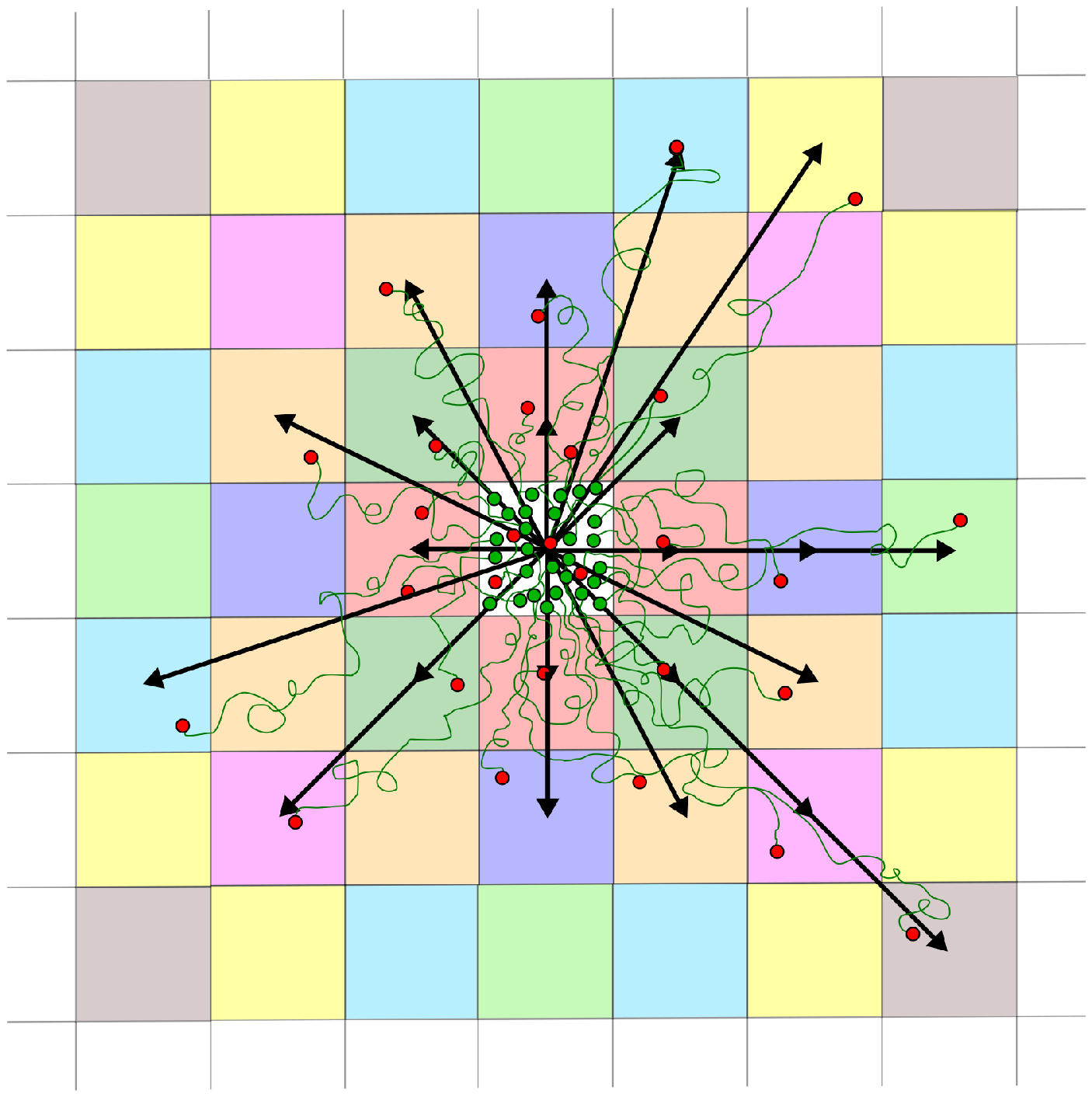}}}
\subfloat[\label{subfig-2:mdlg_full_49}]{{\includegraphics[width=0.45\textwidth]{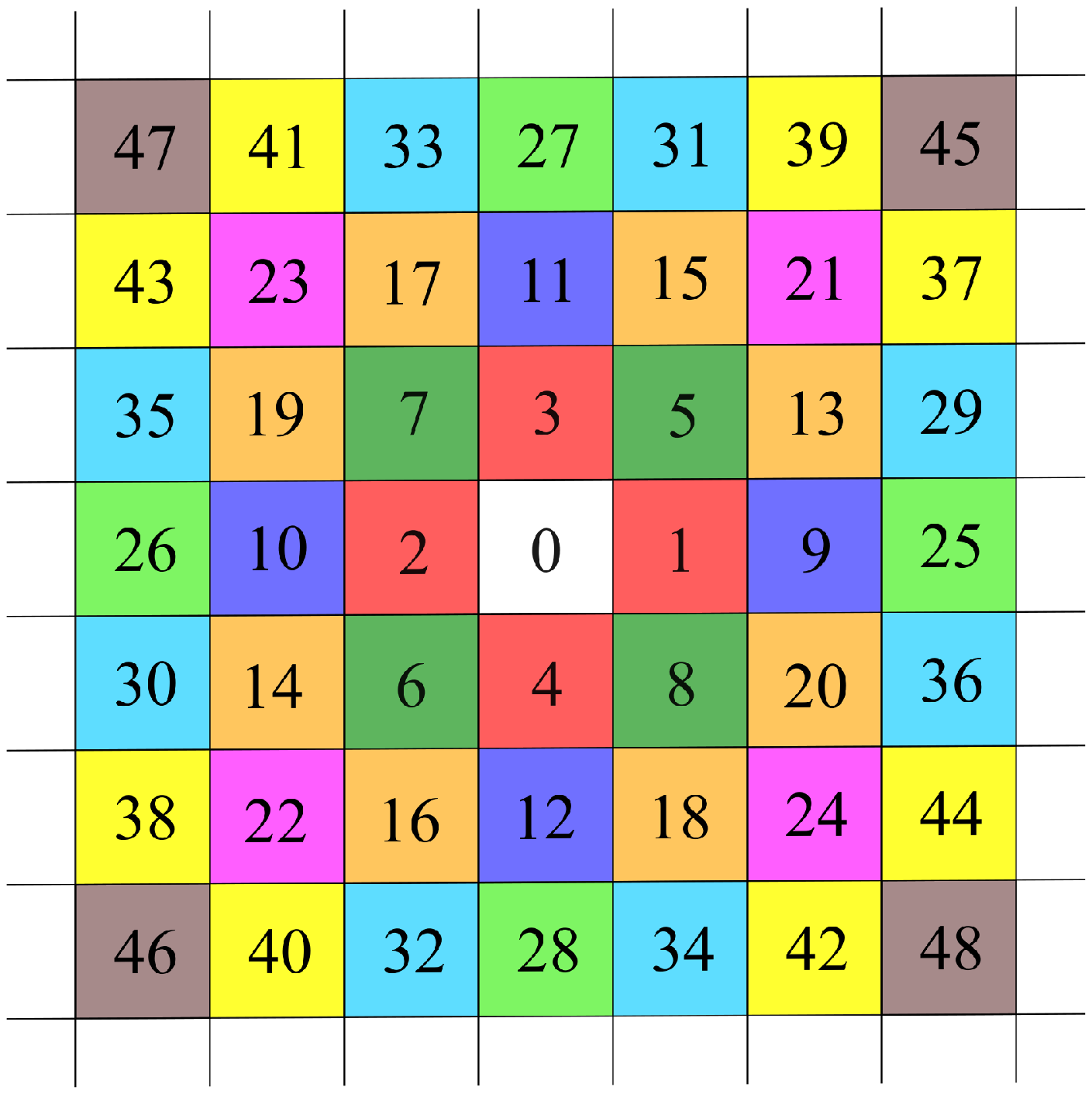}}}
\caption{\textcolor{black}{(Color online) (a) Sketch of the MDLG analysis. A lattice is superimposed onto the MD simulation domain. The movement of the particles is tract from the central node using their MD trajectories. The green circles represent the position of the particles at time $t-\Delta t$ and the red circles are their respective positions at time $t$. Using the particle trajectories and the imposed lattice, the occupation number $n_i$ is defined as given in Eq.\,(\ref{eq:n_i}). The black arrows are the lattice velocities. Only the lattice velocities which have at least one particle within their area (i.e. non\nobreakdash-zero occupation number) are shown. (b) Schematic representation of the D2Q49 lattice with the numbering convention for the lattice velocities in two dimensions. The central point 0 corresponds to the zeroth-velocity $v_0=(0,0)$ and the rest of the velocities are given as a vector connecting the central point and the lattice point in question as shown in (a). The velocities are color coded depending on their length. }}
\label{fig:mdlg_model}
\end{figure}

In the MDLG analysis, we impose a lattice onto an MD simulation of Lennard-Jones particles and track the migration of the particles from one lattice position to another with displacement $v_i$ after a time step $\Delta t$ \textcolor{black}{as shown in Fig.\,\ref{subfig-1:mdlg_dynamics}. A schematic representation of the lattice is given in Fig.\,\ref{subfig-2:mdlg_full_49} where the numbers 0 to 49 represent the $i$ index of the occupation number of an D2Q49 velocity set. We run molecular dynamics simulations and analyze the particles' trajectories to obtain MDLG occupation numbers defined as}
\begin{equation}
    n_i(x,t) = \sum_j\Delta_x[x_j(t)]\Delta_{x-v_i}[x_j(t-\Delta t)],
    \label{eq:n_i}
\end{equation}
with the delta function $\Delta_x[x_j(t)]=1$, if particle $x$ is in the lattice cell at time $t$, and $\Delta_x[x_j(t)]=0$, otherwise. Here, the $x_j(t)$ is the position of the $j$-th particle at time $t$ \textcolor{black}{and $v_i$ is the particle displacement, which in the MDLG description is strongly correlated to the lattice velocities}. \textcolor{black}{We can now cast the evolution of the occupation numbers $n_i$ in the form of a lattice gas evolution equation as}
\begin{equation}
    n_i(x+v_i, t+\Delta t) = n_i(x,t)+\Xi_i,
    \label{eq:mdlg_evolution}
\end{equation}
\textcolor{black}{by defining the lattice gas collision operator $\Xi_i$ in terms of the occupation numbers as}
\begin{equation}
    \Xi_i = n_i(x+v_i, t+\Delta t) - n_i(x,t).
\end{equation}
The molecular dynamics lattice Boltzmann (MDLB) distribution function is defined as a Boltzmann ensemble average of the MDLG occupation numbers $n_i$ and it is given by
\begin{equation}
        f_i = \langle n_i\rangle_\mathrm{neq}.
\end{equation}
By taking the non\nobreakdash-equilibrium ensemble average of Eq.\,(\ref{eq:mdlg_evolution}), we obtain the MDLB evolution equation
\begin{equation}
    f_i(x+v_i, t+\Delta t) = f_i(x,t)+\Omega_i,\qquad \text{with}\quad \Omega_i = \langle \Xi_i\rangle_\mathrm{neq},
\end{equation}
where $\Omega_i$ is the MDLB collision operator. A key element of the LBM is the global equilibrium distribution function, which in MDLB context is defined as an average of the lattice gas densities $n_i$ over the whole MD domain and all iterations of an equilibrium MD simulation. The MDLB equilibrium distribution function is given by
\begin{equation}
\begin{split}
    f_i^\mathrm{eq} &= \langle n_i\rangle_\mathrm{eq} \\
    &=\left\langle \sum_j \Delta_x[x_j(t)]\Delta_{x-v_i}[x_j(t-\Delta t)]\right\rangle_\mathrm{eq} \\
    &= M\int dx_1 \int d\delta x_1\;P^{(1),\mathrm{eq}}(x_1,\delta x_1) \Delta_x[x_1]\Delta_{x-v_i}[x_1-\delta x_1],
    \label{eq:theory_feq}
\end{split}
\end{equation}
where $M$ is the total number of particles and $P^{(1),\mathrm{eq}}$ is the one\nobreakdash-particle displacement distribution function in equilibrium. This allows us to obtain the equilibrium distribution function $f_i^\mathrm{eq}$ analytically from the one\nobreakdash-particle displacements Probability Distribution Function (PDF). 

\section{Derivation of the MDLB equilibrium distribution function}
\label{sec:equilibrium_df}
\textcolor{black}{In the MDLB formulation, the equilibrium distribution function depends solely on the one\nobreakdash-particle displacement distribution function. Thus, knowing $P^{(1),\mathrm
{eq}}$ is crucial for predicting the equilibrium distribution function. In the following subsections, we derive the equilibrium distribution function from (a) a single Gaussian probability distribution function and (b) from a Poisson weighted sum of Gaussians probability distribution function.}
\subsection{Single Gaussian distribution model}
\label{subsec:gaussian_model}
\textcolor{black}{In Parsa \textit{et al.}\,\cite{parsa_lattice_2017} a good approximation of the MDLB equilibrium distribution function is given by a single Gussian in one\nobreakdash-dimension ($d=1$)}
\begin{equation}
    \begin{split}
    P^{\mathrm{G}}_\alpha(\delta x) = \frac{1}{[2\pi\langle(\delta x_\alpha)^2\rangle]^{d/2}}\exp\left[-\frac{(\delta x_\alpha - u_\alpha\Delta t)^2}{2\langle(\delta x_\alpha)^2\rangle}\right],
    \end{split}
    \label{eq:gaussian_df}
\end{equation}
with displacements $\delta x_\alpha$, second order moment $\langle (\delta x_\alpha)^2\rangle$ \textcolor{black}{and mean velocity $u_\alpha$}. The solution factorizes for higher dimensions and it is given by
\begin{equation}
    P^{\mathrm{G}}(\delta x) = \prod_{\alpha=1}^d P^{\mathrm{G}}_\alpha(\delta x).
    \label{eq:p_g_fact}
\end{equation}
Following Eq.\,(\ref{eq:theory_feq}) the equilibrium distribution function can be expressed as
\begin{equation}
    \frac{f^\mathrm{eq,G}_i}{\rho^\mathrm{eq}} = \prod_{\alpha=1}^d f_{i,\alpha}^\mathrm{eq,G},
    \label{eq:feq_prod}
\end{equation}
\textcolor{black}{with $\rho^\mathrm{eq}$ being the mass density. The equilibrium distribution function $f_{i,\alpha}^\mathrm{eq,G}$ in one\nobreakdash-dimension is given by}
\begin{equation}
\begin{split}
    f_{i,\alpha}^\mathrm{eq,G} &= N\left(e^{-\frac{(u_{i,\alpha}-1)^2}{2a^2}}-2e^{-\frac{u_{i,\alpha}^2}{2a^2}}+e^{-\frac{(u_{i,\alpha}+1)^2}{2a^2}}\right)\\
    &+ \frac{u_{i,\alpha}-1}{2}\left[\mathrm{erf}\left(\frac{u_{i,\alpha}-1}{a\sqrt{2}}\right)-\mathrm{erf}\left(\frac{u_{i,\alpha}}{a\sqrt{2}}\right)\right]\\
    &+ \frac{u_{i,\alpha}+1}{2}\left[\mathrm{erf}\left(\frac{u_{i,\alpha}+1}{a\sqrt{2}}\right)-\mathrm{erf}\left(\frac{u_{i,\alpha}}{a\sqrt{2}}\right)\right],
\end{split}
    \label{eq:feq_gaussian}
\end{equation}
with 
\begin{equation}
    a^2 = \frac{\langle(\delta x_\alpha)^2\rangle}{(\Delta x)^2},\quad\qquad N = \frac{a}{\sqrt{2\pi}},\quad\qquad u_{i,\alpha} = v_{i,\alpha} -u_\alpha,
    \label{eq:a2_n_ua}
\end{equation}
\textcolor{black}{where $\langle(\delta x_\alpha)^2\rangle$ is the mean-squared displacement, $\Delta x$ is the lattice size, and $u_\alpha$ is the mean velocity. We have performed MD simulations with mean velocity set to zero, however, we could obtain results for different mean velocities $u_{\alpha}$ by applying a Galilean transformation. We have set the value of $a^2$ to approximately $1/6$ for which the MDLG results agree with the values of the D2Q9 lattice Boltzmann weights.} For details regarding the derivation of the Gaussian equilibrium distribution function, please refer to \cite{parsa_lattice_2017}.

Even though this formulation shows very good agreement with the measured equilibrium distribution function from MD simulations, under more careful investigation we found that the there are discrepancies of up to about 5\% for certain parameter regimes. This means that the displacement distribution function cannot be fully captured by a single Gaussian and a more complex distribution function has to be applied. 

\subsection{Poisson weighted sum of Gaussians model}
\label{subsec:pwsg_model}
In Pachalieva \textit{et al.}\,\cite{pachalieva2020}, we have introduced a correction of the displacements PDF proposed by Parsa \textit{et al.}\,\cite{parsa_lattice_2017} using a Poisson weighted sum of Gaussians (WSG) instead of a single Gaussian distribution function. The Poisson WSG is given by
\begin{equation}
    \begin{split}
    P^{\mathrm{WSG}}(\delta x) = \sum_{c=0}^\infty e^{-\lambda} \frac{\lambda^c}{c!} P^c(\delta x),
    \end{split}
    \label{eq:poisson_v2}
\end{equation}
\textcolor{black}{where the $P^c(\delta x)$ probability distribution function also factorizes for higher dimensions equivalently to the single Gaussian distribution function as given in Eq.\,(\ref{eq:p_g_fact}). The one\nobreakdash-dimensional Poisson weighted sum of Gaussians probability distribution function $P^c_\mathrm{\alpha}(\delta x)$ is then given by}
\begin{equation}
    P^c_\mathrm{\alpha}(\delta x)=\left[\frac{(\lambda+1)}{2\pi (c+1)\langle(\delta x_\alpha)^2\rangle}\right]^{d/2}\exp\left[-\frac{(\lambda+1)(\delta x_\alpha-u_\alpha\Delta t)^2}{2(c+1)\langle (\delta x_\alpha)^2\rangle}\right],
    \label{eq:p_wsg_one}
\end{equation}
where $\delta x_\alpha$ is the displacement in one\nobreakdash-dimension, $\langle (\delta x_\alpha)^2\rangle$ is the second\nobreakdash-order moment, $u_\alpha$ is the mean velocity, $c$ is the number of occurrences, and $\lambda$ is the average number of collisions. \textcolor{black}{The fact that the new displacement distribution function is just a sum of Gaussians makes the calculation of the new MDLG equilibrium functions surprisingly simple. Thus, we obtain
\begin{equation}
    f_i^{eq}=\sum_{c=0}^\infty e^{-\lambda}\frac{\lambda^c}{c!} f_i^{c,eq}.
\end{equation}
The $f_i^{c,eq}$, similar to Eq.\,(\ref{eq:feq_prod}), is given by }
\begin{equation}
    \frac{f^{c,\mathrm{eq}}_i}{\rho^\mathrm{eq}} = \prod_{\alpha=1}^d f_{i,\alpha}^{c,\mathrm{eq}},
\end{equation}
where $\rho^\mathrm{eq}$ is the mass density and $f^{c,\mathrm{eq}}_{i,\alpha}$ in one\nobreakdash-dimension is given by
\begin{equation}
\begin{split}
     f^{c,\mathrm{eq}}_{i,\alpha} = &\left\{\frac{N_c}{2\sqrt{\pi}}\left(e^{-\frac{(u_{i,\alpha}-1)^2}{N_c^2}}-2e^{-\frac{u_{i,\alpha}^2}{N_c^2}}+e^{-\frac{(u_{i,\alpha}+1)^2}{N_c^2}}\right) \right.\\
    & \left. +\frac{(u_{i,\alpha}-1)}{2}\left[\mathrm{erf}\left(\frac{(u_{i,\alpha}-1)}{N_c}\right)-\mathrm{erf}\left(\frac{u_{i,\alpha}}{N_c}\right)\right] \right.\\
    &\left.+\frac{(u_{i,\alpha}+1)}{2}\left[\mathrm{erf}\left(\frac{(u_{i,\alpha}+1)}{N_c}\right)-\mathrm{erf}\left(\frac{u_{i,\alpha}}{N_c}\right)\right]\right\},
\end{split}\label{eq:f_eq_wsg}
\end{equation}
with 
\begin{equation}
    N_c = \sqrt{\frac{2a^2(c+1)}{\lambda+1}}
\end{equation}
\textcolor{black}{where $a^2$ and $u_{i,\alpha}$ are defined in Eq.\,(\ref{eq:a2_n_ua})}. \textcolor{black}{The one\nobreakdash-dimensional equilibrium distribution function given in Eq.\,(\ref{eq:f_eq_wsg}) is similar to the single Gaussian equilibrium distribution function in Eq.\,(\ref{eq:feq_gaussian}), however, their weighting factors are not the same. The equilibrium distribution function Eq.\,(\ref{eq:f_eq_wsg}) takes also into account the average number of collisions $\lambda$, which needs to be defined.}

\textcolor{black}{One way to approximate the average number of collisions $\lambda$ is by using the velocity auto\nobreakdash-correlation function. However, the auto-correlation function is just a theoretical approximation and is not exact. To eliminate the second\nobreakdash-order and the fourth\nobreakdash-order moment errors, we match these moments to the corresponding ones measured directly from the MD simulations. The second\nobreakdash-order moment of the Poisson WSG one\nobreakdash-particle distribution function can be derived from the second\nobreakdash-order Gaussian integral
\begin{equation}
    \begin{split}
        \mu_2 &=\int_{-\infty}^\infty P^\mathrm{WSG}(\delta x)(\delta x)^2\, d\delta x\\ 
        &=  \int_{-\infty}^\infty \sum_{c=0}^\infty e^{-\lambda}\frac{\lambda^c}{c!} \frac{\sqrt{\lambda+1}}{\sqrt{2\pi(c+1)\langle(\delta x)^2\rangle}}\exp\left(-\frac{(\lambda+1)(\delta x-u\Delta t)^2}{2(c+1)\langle(\delta x)^2\rangle}\right) (\delta x)^2\,d\delta x\\
        &= \langle(\delta x)^2\rangle.
    \end{split}
\end{equation}
Analogous, we obtain the fourth\nobreakdash-order moment from the fourth\nobreakdash-order Gaussian integral
\begin{equation}
    \begin{split}
    \mu_4 &= \int_{-\infty}^\infty P^\mathrm{WSG}(\delta x)(\delta x)^4 \,d(\delta x)\\ 
    &= \int_{-\infty}^\infty \sum_{c=0}^\infty e^{-\lambda} \frac{\lambda^c}{c!} \frac{\sqrt{\lambda+1}}{\sqrt{2\pi (c+1)\langle(\delta x)^2\rangle}}\exp\left(-\frac{(\lambda+1)(\delta x-u\Delta t)^2}{2(c+1)\langle(\delta x)^2\rangle}\right) (\delta x)^4 \,d\delta x\\ 
    &=\frac{3 \langle(\delta x)^2\rangle^2}{(\lambda+1)^2}\left[\lambda^2+3\lambda+1\right].
   \end{split}
   \label{eq:fourth_order}
\end{equation}
By solving the quadratic equation for $\lambda$
\begin{equation}
    \frac{3 \mu_2^2}{(\lambda+1)^2}\left[\lambda^2+3\lambda+1\right] -\mu_4 = 0
    \label{eq:lambda_eq}
\end{equation}
we find the following solutions 
\begin{equation}
   \hspace{-3mm}\lambda_{1,2} =\frac{-9\mu_2^2\pm\sqrt{3[15\mu_2^4-4\mu_2^2\mu_4]}+2\mu_4}{2[3\mu_2^2-\mu_4]}.
\end{equation}
where $\mu_2 = \langle(\delta x)^2\rangle$ and $\mu_4$ are the second\nobreakdash- and fourth\nobreakdash-order displacement moments, respectively. We use the moments measured from MD simulations, which ensures that the Poisson weighted sum of Gaussians model has the same $\mu_2$ and $\mu_4$ moments. In Pachalieva \textit{et al.}\,\cite{pachalieva2020}, we show that $\lambda_{2}$ provides an optimal solution, which we use to derive the Poisson WSG equilibrium distribution function. For detailed derivation and discussion of the Poisson WSG displacement distribution function, please refer to Pachalieva \textit{et al.}\,\cite{pachalieva2020}.}

\begin{figure}[!t]
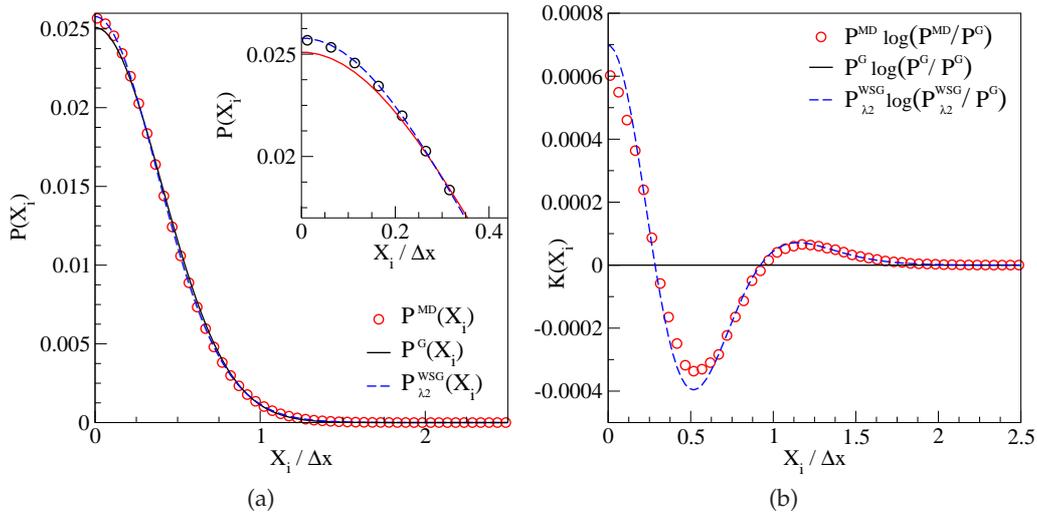

\centering
\subfloat[\label{subfig-1:pdfs}]{{\includegraphics[width=0.49\textwidth]{df_3.2_symbols_dsfd_zoom.eps}}}\hspace{0.18cm}
\subfloat[\label{subfig-1:kld}]{{\includegraphics[width=0.49\textwidth]{kld_1_3.2_symbols_dsfd.eps}}}
\caption{(Color online) (a) Displacements probability distribution functions. The symbols (red) depict a PDF obtained from an MD simulation of LJ particles in equilibrium. The line (black) illustrates a Gaussian probability distribution function defined in Eq.\,(\ref{eq:gaussian_df}) with mean\nobreakdash-squared displacement fitted directly to the MD data. The dashed line (blue) represents the Poisson WSG obtained from Eq.\,(\ref{eq:poisson_v2}). Only the data for positive velocities has been depicted due to symmetry. (b) shows the difference between the distributions per interval $X_i$ as defined in Eq.\,(\ref{eq:k_Xi}). The presented data is for the standard parameters used in the paper and a coarse\nobreakdash-grained time step $\Delta t = 3.2$.}
\label{fig:df_kld}
\end{figure}

\textcolor{black}{Meaningfully comparing two probability distribution functions is a non-trivial task since often there are significant deviations in the tails of the distribution that would show up in a simpler measure like dividing the distributions. However, since the tails carry little weight, these deviations are not relevant for the system. In Pachalieva \textit{et al.} \cite{pachalieva2020}, we used the Kullback-Leibler (KL) divergence, a tool commonly used in machine learning. The element\nobreakdash-wise definition of this function is given by}
\begin{equation}
     K(X_i) = K(R\parallel Q) = R(X_i)\log \left({\frac {R(X_i)}{Q(X_i)}}\right),
     \label{eq:k_Xi}
\end{equation}
where $R(X_i)$ and $Q(X_i)$ are probability distributions over an interval $X_i$. By performing a sum over all the bins $X_i$, we obtain the Kullback\nobreakdash-Leibler (KL) divergence \cite{kullback1951information} defined as 
\begin{equation}
     D_{\text{KL}}(R\parallel Q)=\sum_{i}R(X_i)\log \left({\frac {R(X_i)}{Q(X_i)}}\right).
     \label{eq:kullback_leibler}
\end{equation}
The KL divergence measures the discrepancies of one probability distribution function to another. It is always non\nobreakdash-negative $D_{\text{KL}}(R\parallel Q)\geq0$ or equal to zero if and only if the probability distribution functions are identical $R(X_i) = Q(X_i)$. 

In Fig.\,\ref{subfig-1:pdfs}, we see the true probability distribution function obtained from the MD data $P^\mathrm{MD}(X_i)$, the Gaussian probability distribution function $P^\mathrm{G}(X_i)$, and the Poisson WSG distribution function $P^\mathrm{WSG}(X_i)$. There is a visible divergence between the Gaussian and the other two distribution functions. \textcolor{black}{We measured the element\nobreakdash-wise Kullback\nobreakdash-Leibler divergence $K(X_i)$, as defined in Eq.\,(\ref{eq:k_Xi}),} for $P^\mathrm{G}(X_i)$ compared to the MD data and the Poisson WSG distribution function as shown in Fig.\,\ref{subfig-1:kld}. The results suggest that even though the Gaussian and the Poisson WSG probability distribution functions have the same second moment, their deviations in \textcolor{black}{the fourth\nobreakdash- and higher\nobreakdash-order} moments influence strongly the form of the distribution function. In Section\,\ref{sec:results}, we show how these deviations effect the LBM equilibrium distribution function.

\section{Simulations setup}
\label{sec:sim_setup}
\textcolor{black}{All measured data, from probability distribution functions of the displacements $P^\mathrm{MD}(X_i)$ to the equilibrium distribution function $f_i^\mathrm{eq,MD}$ depicted in  Figs.\,\ref{fig:feq_not_scaled}-\,\ref{fig:feq_n9-25}, are obtained from molecular dynamics simulations.} To perform the MD simulations we used the open\nobreakdash-source molecular dynamics framework LAMMPS \cite{plimpton_fast_1995, noauthor_lammps_nodate} developed by Sandia National Laboratories. \textcolor{black}{The LAMMPS package uses Velocity\nobreakdash-Verlet integration scheme.} The MD simulations consist of particles interacting with the standard 6\nobreakdash-12 Lennard\nobreakdash-Jones (LJ) intermolecular potential given by
\begin{equation}
    V_{LJ} = 4\varepsilon\left[\left(\frac{\sigma}{r}\right)^{12}-\left(\frac{\sigma}{r}\right)^{6}\right],
\end{equation}
with $\sigma$ being the distance at which the inter\nobreakdash-particle potential goes to zero, $r$ is the distance between two particles, and $\varepsilon$ is the potential well depth. The particle mass and the LJ particle diameter are set to $m=1$ and $\sigma=1$, respectively. \textcolor{black}{The LJ timescale is given by the time needed for a particle with kinetic energy of half the potential energy well $\varepsilon$ to traverse one diameter $\sigma$ of an LJ particle. This can be also expressed as 
\begin{equation}
    \tau_{\mathrm{LJ}} = \sqrt{\frac{m\sigma^2}{\varepsilon}}.
\end{equation}
The thermal time scale corresponds to the time it takes a particle with the kinetic energy of $1/2\;k_BT$ to transverse the diameter $\sigma$ of a LJ particle, which is given by 
 \begin{equation}
     \tau_\mathrm{th}=\sqrt{\frac{m\sigma^2}{k_B T}}.
 \end{equation}
We  executed molecular dynamics simulations with temperature of $20$ in the LJ units defined above. This corresponds to a thermal time scale smaller than the LJ time scale $\tau_\mathrm{LJ}$ by factor of $1/\sqrt{20}\approx 0.22$.}

The number of particles in each simulation has been fixed to $N=99\ 856$ which fills a two\nobreakdash-dimensional (2D) square with length L = 1000$\sigma$. The area fraction $\phi$ of the domain is calculated from the area of the circular LJ particles multiplied by the number of particles divided by the area of the domain, where the diameter of the circular LJ particle is given by $\sigma$. The MD simulations considered in this publication have an area fraction of $\phi = 0.078387$. We initialised the simulations using homogeneously distributed particles with kinetic energy corresponding to temperature equal to 20 in LJ units. \textcolor{black}{This corresponds to a dilute gas with high temperature. The temperature is way above the critical temperature for liquid-gas coexistence of $T_c=1.3120(7)$, and the density is way below the critical density $\rho_c=0.316(1)$ \cite{potoff1998}. We focus our attention to MD simulations of a fairly dilute gas in equilibrium, since the assumption that the collision times is Poisson distributed is correct only for dilute systems.}

\begin{table}[!t]
    \centering
		\caption{\textcolor{black}{Initialization parameters of the molecular dynamics simulations performed using LAMMPS framework. For all MD simulations the MD step size is fixed to $0.0001\tau_{\mathrm{LJ}}$ and the number of coarse-grained iterations is $2\,000$.}}
        \begin{tabular}{ c c r r r r }
        \hline
           &           &     & MD output            &Total MD\\
$\Delta t$ &$\Delta x$ &$lx$ & frequency            &time    \\
           & & & ($1/\tau_{\mathrm{LJ}}$)   &($\tau_{\mathrm{LJ}}$)    \\
        \hline
        0.3911 & 4       &250   &   3\,911 &    782.2  \\
        0.5000 & 5       &200   &   5\,000 & 1\,000.0  \\
        0.5626 & 5.5     &180   &   5\,626 & 1\,125.2  \\
        0.6927 & 6.6(6)  &150   &   6\,927 & 1\,385.4  \\
        0.9009 & 8.3(3)  &120   &   9\,009 & 1\,801.8  \\
        1.1261 &10       &100   &11\,261   & 2\,252.2  \\
        1.4994 &12.5     &80    &14\,994   & 2\,998.8  \\
        1.6342 &13.3(3)  &75    &16\,342   & 3\,268.4  \\
        2.0338 &15.625   &64    &20\,338   & 4\,067.6  \\
        2.9280 &20       &50    &29\,280   & 5\,856.0  \\
        4.1821 &25       &40    &41\,821   & 8\,364.2  \\
        6.1751 &31.25    &32    &61\,751   &12\,350.2  \\\hline
        \label{tab:simulation_setup}
        \end{tabular}
\end{table}

\textcolor{black}{Since the MD simulations correspond to a dilute high temperature gas, the particle velocities will be also larger than for a typical molecular dynamics simulation. Thus, we set the MD step size is to $0.0001\,\tau_{\mathrm{LJ}}$ which is considerably small to ensure high accuracy of the MD data.} We define a dimensionless coarse\nobreakdash-grained time step $\Delta t$ being the product of the MD step size and the MD output frequency shown in Table\,\ref{tab:simulation_setup}. The time step $\Delta t$ is chosen such that the MD simulations are restricted to the ratio of the mean\nobreakdash-squared displacement and the squared lattice size being set to
\begin{equation}
    a^2 = \frac{\langle(\delta x)^2\rangle}{(\Delta x)^2} \approx 0.1611,
    \label{eq:a2}
\end{equation}
\textcolor{black}{this corresponds to the parameter $a^2$ given in Eq.\,(\ref{eq:a2_n_ua}), which has been also used in earlier publications \cite{parsa_lattice_2017,parsa_validity_2019}. By fixing the value, we ensure that most of the LJ particles in equilibrium will travel up to one lattice space which corresponds to an D2Q9 lattice Boltzmann method.} To verify that the Poisson WSG equilibrium distribution function $f_i^{\mathrm{eq,WSG}}$ approximated the MD data better than the single Gaussian equilibrium distribution function $f_i^{\mathrm{eq,G}}$ across the length scale, from ballistic to diffusive regime, we vary the coarse\nobreakdash-grained time step $\Delta t \in [0.3911, 6.1751]$ and the lattice size $\Delta x \in [4, 31.25]$ of the executed simulations. An overview of the MD simulation setup is given in Table\,\ref{tab:simulation_setup}. The number of lattice points $lx$ varies from 250 to 32 depending on the coarse\nobreakdash-grained time step $\Delta t$. For each coarse\nobreakdash-grained time step $\Delta t$ we performed $2\,000$ iterations which corresponds to total MD time of $782.2\,\tau_{\mathrm{LJ}}$ to $12\,350.2\,\tau_{\mathrm{LJ}}$ for the smallest and largest coarse\nobreakdash-grained time step $\Delta t$, respectively. In order to bring the molecular dynamics simulations to equilibrium state before we start collecting data, the initial 3\,000\,000 iterations of each simulation were discarded. The discarded iterations are not included in Table\,\ref{tab:simulation_setup} for clarity.

The MD simulation setup characterizes a hot dilute gas in equilibrium with average velocity \textcolor{black}{$u_\alpha$} fixed to zero
\begin{equation}
    Nu_\alpha = \sum_{j=1}^Nv_{j,\alpha} = 0,
\end{equation}
where N is the number of LJ particles.

We performed standard molecular dynamics simulations without thermostat. In the LAMMPS framework this is called NVE integration. The microcanonical ensamble NVE is characterized by constant number of particles (N), constant volume (V) and constant energy (E).

\section{Results}
\label{sec:results}
In order to obtain a measured equilibrium distribution function, we post\nobreakdash-process the collected MD data using the MDLG analysis tool. The MD domain is overlapped with a lattice and we trace the migration of the particles over time from one lattice to another. By doing this, we obtain the MDLG occupation numbers $n_i(x,t)$ as defined in Eq.\,(\ref{eq:n_i}) which after sufficient averaging deliver the MDLB equilibrium distribution function $f_i^{\mathrm{eq, MD}}$as defined in Eq\,(\ref{eq:theory_feq}).

The analytical models of the equilibrium distribution function defined in Section\,\ref{sec:equilibrium_df} depend only on the choice of the one\nobreakdash-particle displacement distribution function. Since we define two different one\nobreakdash-particle distribution, we expect to see also changes in the respective equilibrium distribution function derived from them, even though their second\nobreakdash-order moments are equivalent. However, a non\nobreakdash-trivial question remains how the migration of particles from one node to another changes within a lattice. 

\begin{figure}[!h]
\centering
\subfloat[\label{subfig-1:feq_not_scaled}]{{\includegraphics[width=0.74\textwidth]{feq_pwsg_g_paper_reduced_ledgend.eps}}}\hspace{-1.2cm}
  \subfloat[\label{subfig-1:mdlg_all}]{{\includegraphics[width=0.28\textwidth]{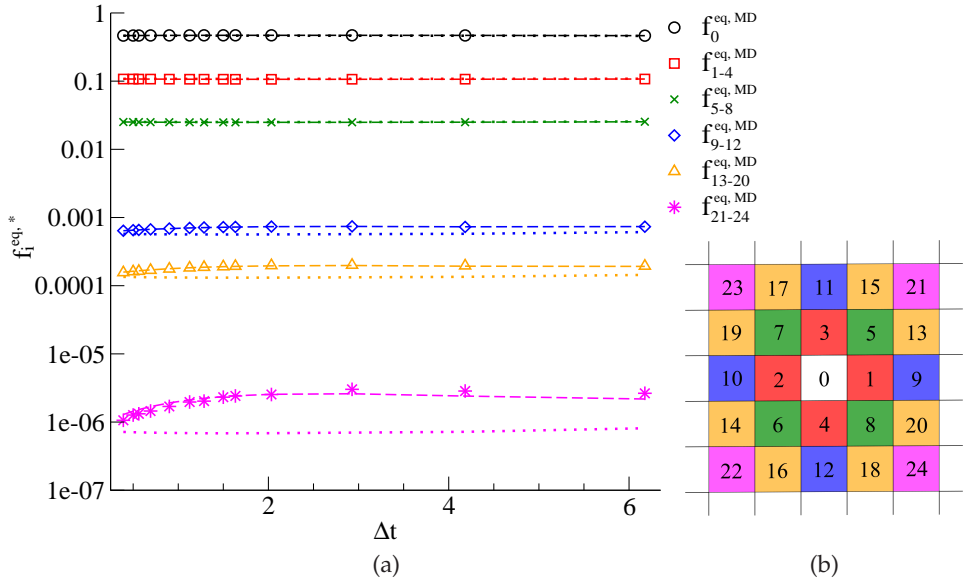}}}
\caption{(Color online) (a) Estimated equilibrium distribution functions $f_i^{\mathrm{eq,*}}$ obtained either from MD simulation data ($f_i^{\mathrm{eq,MD}}$) depicted with symbols, theoretical solution using a single Gaussian distribution function ($f_i^{\mathrm{eq,G}}$) depicted with dotted lines or theoretical solution using Poisson WSG ($f_i^{\mathrm{eq,WSG}}$) depicted with dashed lines. (b) Our numbering for the velocities in a D2Q25 lattice. The equilibrium distribution function $f_i^{\mathrm{eq,*}}$ values are color coded and each color represents one of the six sets of equilibrium distribution function contributions. Here, the asterisk ($^*$) corresponds to the variety of methods used to obtain an equilibrium distribution function: measured from MD simulation, single Gaussian analytical solution and Poisson WSG analytical solution. \textcolor{black}{Note that by using a simple-minded direct comparison on a log-scale (rather than a Kulbeck-Leibler measure) practically irrelevant errors for very small occupation numbers stand out here.}}
\label{fig:feq_not_scaled}
\end{figure}

To gain a better understanding, we calculate the equilibrium distribution function for an extended D2Q25 lattice which corresponds to two neighboring cells in $X$- and $Y$-directions for a two\nobreakdash-dimensional domain. A schematic representation of the D2Q25 lattice is given in Fig.\,\ref{subfig-1:mdlg_all}. In equilibrium state with zero initial velocity, one distinguishes six sets of equilibrium distribution function contributions: $f_{0}^\mathrm{eq,*}, f_{1-4}^\mathrm{eq,*}, f_{5-8}^\mathrm{eq,*}, f_{9-12}^\mathrm{eq,*}, f_{13-20}^\mathrm{eq,*},$ and $f_{21-24}^\mathrm{eq,*},$ where each set has a unique displacement length from the central lattice. When measuring the equilibrium distribution function $f_{i}^\mathrm{eq,MD}$ from the MD simulations, we average over the number of lattices for each set to obtain a symmetric probability distribution function. It is worth mentioning that the deviations of the $f_{i}^{\mathrm{eq,MD}}$ values within each set are relatively small. 

The MDLG analysis was introduced for an D2Q49 lattice including a third layer of neighbouring cells, however, the number of considered neighboring layers depends solely on the problem at hand. For a simulation in equilibrium with zero velocity, and a parameter $a^2$ as defined in Eq.\,(\ref{eq:a2}) being set to approximately $0.1611$, we obtain an equilibrium distribution function which is symmetric and has significant contributions up to D2Q25 lattice nodes. 

The estimated equilibrium distribution function $f_i^{\mathrm{eq,*}}$ for a variety of coarse\nobreakdash-grained time steps $\Delta t \in [0.3911, 6.1751]$ is shown in Fig.\,\ref{subfig-1:feq_not_scaled}. The equilibrium distribution function $f_i^{\mathrm{eq,*}}$, as mentioned above, is obtained from three different methods: $f_i^{\mathrm{eq,MD}}$ is measured from an MD simulation, $f_i^{\mathrm{eq,G}}$ is theoretically estimated using a single Gaussian distribution function and $f_i^{\mathrm{eq,WSG}}$ is theoretically estimated from a Poisson WSG distribution function. The theoretical equilibrium distribution function models are described in detail in Sections \ref{sec:equilibrium_df}\,\ref{subsec:gaussian_model} and \ref{sec:equilibrium_df}\,\ref{subsec:pwsg_model}, respectively. 

\begin{figure}[!t]
  \centering
  \subfloat[\label{subfig-1:feq_n0-8}]{{\includegraphics[width=0.86\textwidth]{feq_pwsg_g_paper_scaled_G_n0-8.eps}}}\hspace{-2.8cm}
  \subfloat[\label{subfig-1:mdlg_n0-8}]{{\includegraphics[width=0.28\textwidth]{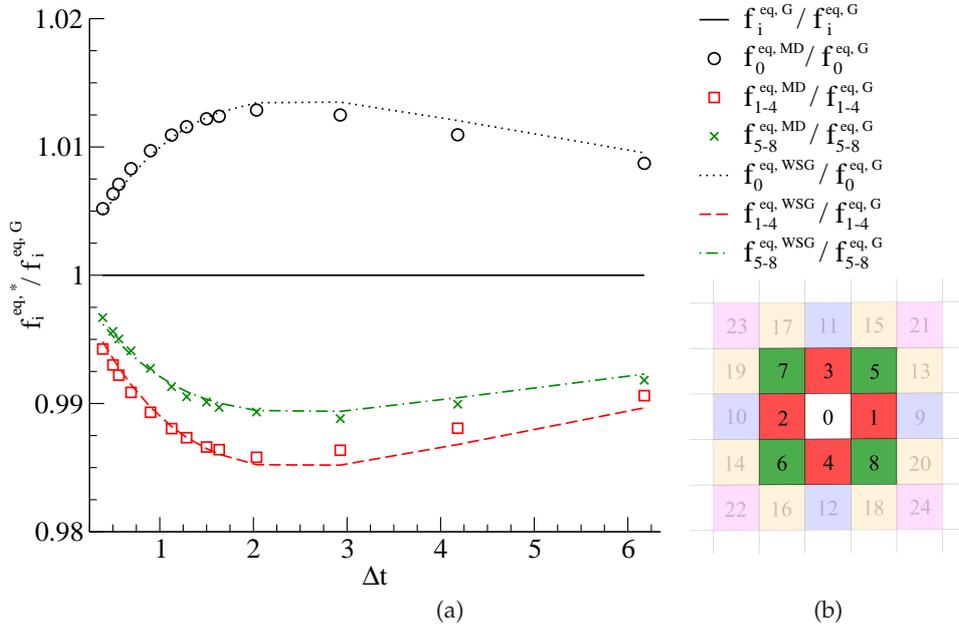}}}
  \caption{(Color online) (a) First layer equilibrium distribution functions $f_{0-8}^{\mathrm{eq,*}}$ scaled to the Gaussian equilibrium distribution function.  The equilibrium distribution functions are obtained either from MD simulation data ($f_{0-8}^{\mathrm{eq,MD}}$), theoretical solution using a single Gaussian distribution function ($f_{0-8}^{\mathrm{eq,G}}$) or theoretical solution using Poisson WSG ($f_{0-8}^{\mathrm{eq,WSG}}$). (b) Schematic representation of the D2Q25 lattice. The equilibrium distribution function $f_i^{\mathrm{eq,*}}$ values are color coded and each color represents one of the six sets of equilibrium distribution function contributions. Here, the asterisk ($^*$) corresponds to the variety of methods used to obtain an equilibrium distribution function: measured from MD simulation, single Gaussian analytical solution and Poisson WSG analytical solution.}
\end{figure}

In Fig.\,\ref{subfig-1:feq_not_scaled} one can see that the largest equilibrium distribution function contributions are coming from the first layer neighbours $f_{0-8}^{\mathrm{eq,*}}$. These nodes are approximated very well by both theoretical models, please refer to Fig.\,\ref{subfig-1:feq_n0-8} for a detailed comparison of the measured and the theoretical $f_{0-8}^{\mathrm{eq,*}}$. The next equilibrium distribution function groups $f_{9-12}^{\mathrm{eq,*}}$ and $f_{13-20}^{\mathrm{eq,*}}$ are significantly smaller than $f_{0-8}^{\mathrm{eq,*}}$ with one to two order of magnitude. For $f_{9-12}^{\mathrm{eq,*}}$ and $f_{13-20}^{\mathrm{eq,*}}$, we see that the deviations of the measured and the theoretical single Gaussian model become larger. The Poisson WSG $f_{9-20}^{\mathrm{eq,*}}$ show a very good agreement with the measured equilibrium distribution function. The diagonal nodes in the second layer $f_{21-25}^{\mathrm{eq,*}}$ are even smaller and their value could be considered negligible. However, the measured equilibrium distribution function $f_{i}^{\mathrm{eq,MD}}$ shows a good agreement with the theoretical Poisson WSG $f_{i}^{\mathrm{eq,WSG}}$ even for very small contributions such as $f_{21-25}^{\mathrm{eq,*}}$. This suggests that these contributions even though really small are not just noise but theoretically justified.

\begin{figure}[!h]
  \centering
  \subfloat[\label{subfig-1:feq_n9-25}]{{\includegraphics[width=0.86\textwidth]{feq_pwsg_g_paper_scaled_G_n9-25.eps}}}\hspace{-2.8cm}
  \subfloat[\label{subfig-1:mdlg_n9-25}]{{\includegraphics[width=0.28\textwidth]{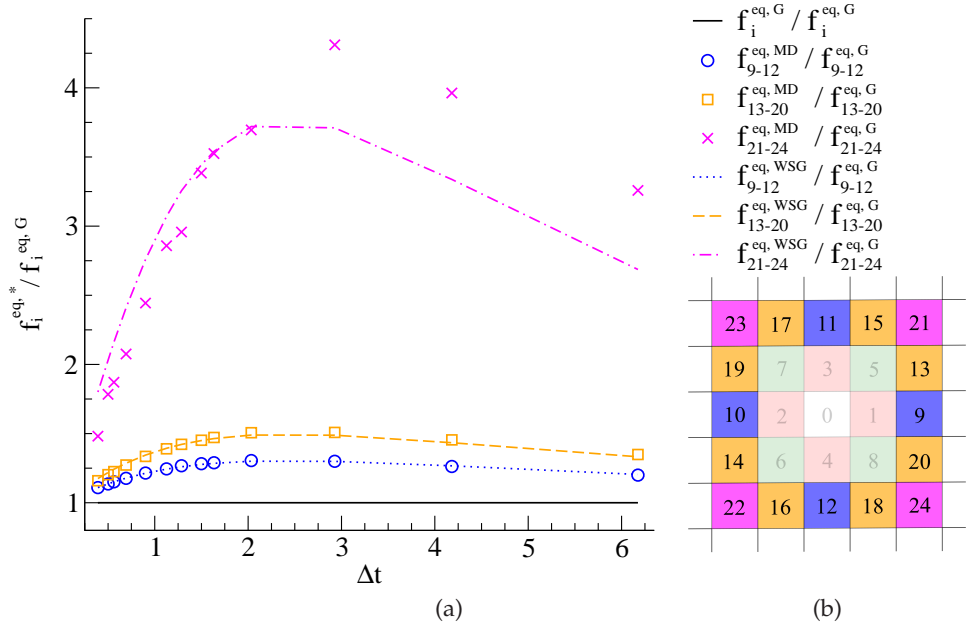}}}
  \caption{(Color online) (a) Second layer equilibrium distribution functions $f_{9-24}^{\mathrm{eq,*}}$ scaled to the Gaussian equilibrium distribution function. The equilibrium distribution functions are obtained either from MD simulation data ($f_{9-24}^{\mathrm{eq,MD}}$), theoretical solution using a single Gaussian distribution function ($f_{9-24}^{\mathrm{eq,G}}$) or theoretical solution using Poisson WSG ($f_{9-24}^{\mathrm{eq,WSG}}$). (b) Schematic representation of the D2Q25 lattice. The equilibrium distribution function $f_i^{\mathrm{eq,*}}$ values are color coded and each color represents one of the six sets of equilibrium distribution function contributions. Here, the asterisk ($^*$) corresponds to the variety of methods used to obtain an equilibrium distribution function: measured from MD simulation, single Gaussian analytical solution and Poisson WSG analytical solution.}
  \label{fig:feq_n9-25}
\end{figure}

Figures \ref{subfig-1:feq_n0-8} and \ref{subfig-1:feq_n9-25} depict the equilibrium distribution functions scaled to the single Gaussian equilibrium function. They show how the measured from MD simulation and the novel Poisson WSG equilibrium distribution functions deviate from the single Gaussian. The first layer equilibrium distribution function values are shown in Fig.\,\ref{subfig-1:feq_n0-8}. These nodes have the largest contribution to the total equilibrium distribution function. 

Fig.\,\ref{subfig-1:feq_n0-8} shows more particles staying at node zero and a depression for the first neighbouring layer (nodes 1 to 8). This very same feature repeats itself in Fig.\,\ref{subfig-1:kld}. The $\mathrm{P}_{\lambda 2}^\mathrm{WSG}\mathrm{log}(\mathrm{P}_{\lambda 2}^\mathrm{WSG}/\mathrm{P}^\mathrm{G})$ values depicted in blue, show that the number of small displacements is enhanced $X_i/\Delta x \in [0, 0.3]$ while the number of $X_i/\Delta x \in [0.3, 1.0]$ displacements is suppressed. 

The second layer equilibrium distribution function values are depicted in Fig.\,\ref{subfig-1:feq_n9-25}. \textcolor{black}{As one can see in Fig.\,\ref{subfig-1:kld}, there is an} enhanced probability of large displacements $X_i/\Delta x \in [0.9, 1.6]$ which corresponds to the larger values of $f_{9-24}^{\mathrm{eq,WSG}}$ in Fig.\,\ref{subfig-1:feq_n9-25}. The deviations (up to approx. 4.5\%) from the theoretical single Gaussian equilibrium distribution function are also larger compared to the first layer nodes $f_{0-8}^{\mathrm{eq,WSG}}$. Since the $f_{9-24}^{\mathrm{eq,WSG}}$ true values are smaller by multiple orders of magnitude than the first layer neighbours $f_{0-8}^{\mathrm{eq,WSG}}$ these deviations \textcolor{black}{are almost irrelevant for} the total equilibrium distribution function, even though they are larger. Nevertheless, Fig.\,\ref{subfig-1:feq_n9-25} shows clearly that the Poisson WSG equilibrium distribution function captures \textcolor{black}{the MD data more precisely}. 

\section{Outlook}
\label{sec:outlook}
\textcolor{black}{In this article, we have derived a better approximation for the MDLG equilibrium distribution function. It deviates from the previous best approximation by Parsa \textit{et al.} \cite{parsa_lattice_2017} in a broad transition region between the ballistic and diffusive regime of random particle displacements.}

\textcolor{black}{Despite the fact that these deviations are small, we expect them to be of great importance in the analysis of non\nobreakdash-equilibrium systems, particularly systems not too far from equilibrium, as is typical in hydrodynamic systems. What we have outlined here is the equilibrium behavior of the MDLG mapping of a molecular dynamics simulation onto a lattice gas. The key interest, however, lies in the non\nobreakdash-equilibrium predictions of this mapping. In future research, we will investigate MDLG predictions for lattice gas and lattice Boltzmann collision operators. In such systems we expect to find only small deviations from local equilibrium, and to quantify these small deviations it is essential to have a very good understanding of the equilibrium behavior of the MDLG mapping.}

\enlargethispage{20pt}


\dataccess{This manuscript has no further supporting data.}

\aucontribute{AW supervised the research, contributed to it, and  revised  the  manuscript. AP contributed to the research, embedded the proposed model, set up the test cases, performed the data analysis and wrote the manuscript. All authors read and approved the manuscript.}
\competing{The authors declare that they have no competing interests.}

\funding{AP is partially supported by the Center for Nonlinear Studies (CNLS) and the Laboratory Directed Research and Development (LDRD) program at Los Alamos National Laboratory (LANL), and the German Federal Ministry of Education and Research (BMBF) in the scope of the project Aerotherm (reference numbers: 01IS16016A-B).}



\bibliography{references_05_06}

\end{document}